\documentclass[aps,prl,notitlepage,amsmath,amssymb,superscriptaddress,10pt,nobibnotes]{revtex4-1}
\pdfoutput=1

\usepackage{amsfonts, amssymb, amsmath, dsfont,mathrsfs}
\usepackage{graphicx}
\usepackage{color}
\usepackage[utf8]{inputenc}
\usepackage[T1]{fontenc}
\usepackage[svgnames]{xcolor}
\usepackage{textcomp}
\usepackage{siunitx}

\usepackage{hyperref}
\hypersetup{colorlinks,urlcolor=DarkBlue,linkcolor=DarkBlue,citecolor=DarkBlue,pdfdisplaydoctitle=true,pdfpagemode=UseOutlines,bookmarksnumbered=true,bookmarksopen=true} 

\makeatletter
\renewcommand{\eqref}[1]{\textup{\tagform@{\ref*{#1}}}}
\makeatother

\makeatletter
\renewcommand*{\fnum@figure}{}
\renewcommand*{\@caption@fignum@sep}{}
\makeatother

\newcommand{\fig}[1]{Fig.~\ref{#1}}

\renewcommand{\bm}[1]{\boldsymbol{\mathbf{#1}}}

\newcommand{\beginsupplement}{%
        \setcounter{table}{0}
        \renewcommand{\thetable}{S\arabic{table}}%
        \setcounter{figure}{0}
        \renewcommand{\thefigure}{S\arabic{figure}}%
        \setcounter{section}{0}
        \renewcommand{\thesection}{\Roman{section}}
        \setcounter{equation}{0}
        \renewcommand{\theequation}{S\arabic{equation}}
     }

\begin{document}

\title{Efficient and flexible approach to ptychography using an optimization framework based on automatic differentiation}
\author{Jacob Seifert}
\altaffiliation{\href{mailto:j.seifert@uu.nl}{j.seifert@uu.nl}}
\affiliation{Debye Institute for Nanomaterials Science, Utrecht University, P.O. Box 80000, 3508 TA Utrecht, The Netherlands}
\author{Dorian Bouchet}
\affiliation{Debye Institute for Nanomaterials Science, Utrecht University, P.O. Box 80000, 3508 TA Utrecht, The Netherlands}
\author{Lars Loetgering}
\affiliation{Advanced Research Center for Nanolithography, Science Park 106, 1098 XG Amsterdam, The Netherlands}
\author{Allard P.\ Mosk}
\affiliation{Debye Institute for Nanomaterials Science, Utrecht University, P.O. Box 80000, 3508 TA Utrecht, The Netherlands}

\begin{abstract}
Ptychography is a lensless imaging method that allows for wavefront sensing and phase-sensitive microscopy from a set of diffraction patterns.
Recently, it has been shown that the optimization task in ptychography can be achieved via automatic differentiation (AD). 
Here, we propose an open-access AD-based framework implemented with TensorFlow, a popular machine learning library.
Using simulations, we show that our AD-based framework performs comparably to a state-of-the-art implementation of the momentum-accelerated ptychographic iterative engine (mPIE) in terms of reconstruction speed and quality.
AD-based approaches provide great flexibility, as we demonstrate by setting the reconstruction distance as a trainable parameter.
Lastly, we experimentally demonstrate that our framework faithfully reconstructs a biological specimen.
\end{abstract}

\maketitle

\section{Introduction}

Ptychography is a computational imaging method that allows to recover both the complex illumination and the transmission function of an object~\cite{Rodenburg2019-kw}. The method is based on the illumination of an object with a localized, coherent probe and the measurement of the resulting diffraction patterns. By laterally scanning the object with overlapping illumination areas, the lost phase information in the diffraction intensities can be recovered through phase retrieval. 
Further algorithmic extensions to ptychography and related methods have been introduced. For example, some algorithms allow to weaken the coherence requirement of the probe beam \cite{Thibault2013-kg}, 
to correct for experimental errors in the positioning and the setup geometry \cite{Hurst2010-td, Maiden2012-xo, Dwivedi2018-yy, Loetgering2018-wf}, to obtain three-dimensional reconstructions \cite{Maiden2012-ja, Tsai2016-vi, Tian2015-yj, Lim2019-kv}, or to acquire images without any moving parts using Fourier ptychography~\cite{Horstmeyer2016-gm, Konda2020-jk, Chen2018-mc} or single-shot ptychography~\cite{Chen2018-mc, Sidorenko2016-uw}. 
Conceptionally similar algorithms have been employed for wide-field fluorescence imaging \cite{Yilmaz2015-wz}.
Potential applications of quantitative phase imaging include label-free imaging in biomedicine~\cite{Park2018-mw}, characterization of lens aberrations~\cite{Du2020-pd}, and x-ray crystallography~\cite{Miao2015-cb}.
Many phase retrieval algorithms are based on the alternating projections scheme \cite{Fienup1978-nl}, where an object is iteratively reconstructed from the modulus of its Fourier transform and by applying known constraints. One variant of this approach for ptychography is called the ptychographical iterative engine (PIE) \cite{Rodenburg2004-oi}.
Significant improvements to the convergence and robustness of PIE have been made by enabling the simultaneous reconstruction of the probe beam \cite{Maiden2009-pn} (known as ePIE), and later by revising the update functions and by borrowing the idea of momentum from the machine learning community \cite{Maiden2017-tj}. The latter introduction to the PIE family has been coined momentum-accelerated PIE (mPIE) and can be considered as a state-of-the-art approach to perform ptychographic reconstructions. 


Instead of solving the phase problem in ptychography using algorithms associated with PIE, it has recently been shown that the object can also be recovered by using optimizers based on automatic differentiation (AD) \cite{Nashed2017-rf, Ghosh2018-jf, Kandel2019-da, Du2020-sh}.
This approach allows to solve the inverse problem in ptychography without the need to find an analytical derivation of the update function, which can be challenging to obtain. 
Moreover, it directly benefits from the fast progress in the machine-learning community in terms of software tools, hardware, and algorithms. 
However, it remains unclear how this approach compares to state-of-the-art algorithms, notably in terms of computational time and reconstruction quality.

In this paper, we implement an AD-based reconstruction framework using TensorFlow \cite{Abadi2016-ia} and Keras \cite{Chollet2015-oo} and benchmark it against an implementation of mPIE running on a graphics processing unit (GPU)~\cite{Loetgering2018-wf}.
Keras enables us to solve the reconstruction problem in ptychography using a layer-by-layer approach, akin to the architecture of deep neural networks. As a result, our framework is highly modular, and it is straightforward to extend the physics model or implement specific cost functions, which is not always possible using PIE. Lastly, the layered architecture makes our AD-based framework interesting for the emergent idea of interfacing a deep neural network (DNN) directly to the forward model~\cite{kellman2019physicsbased, Barbastathis2019-ne}. We are making the source code for our AD-based framework available in code file~1~\cite{seifert2020}. Thus, we provide a flexible tool to perform ptychographic reconstructions and simultaneously estimate unknown experimental parameters, opening interesting perspectives to improve the reconstruction quality for visible and X-ray lensless imaging techniques.

\section{Problem formulation}

\begin{figure}[ht]
\centering
\includegraphics[width=11.5cm]{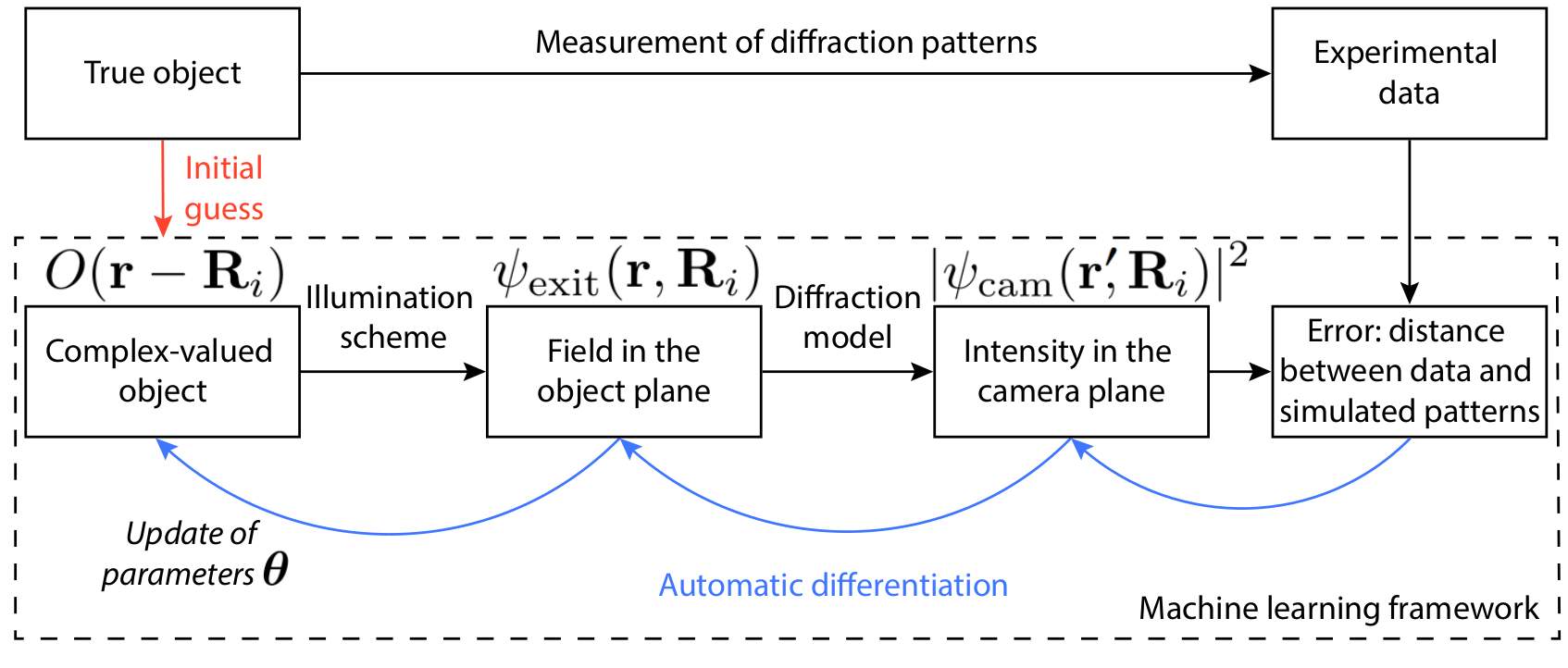}
\caption{Flowchart of the optimization framework based on automatic differentiation. The dataflow from left to right represents the forward model in ptychography. The object is initially guessed and adapted by the optimizer in order to minimize the error between the experimental and predicted diffraction patterns. By applying the chain rule, automatic differentiation finds the derivative of the error function with respect to the independent variables (such as object, probe, or parameters of the setup geometry). Gradient descent optimization then backpropagates the changes to the variables.}
\label{flowchart}
\end{figure}

The AD-based reconstruction framework implemented in this paper is presented in \fig{flowchart}. Let $O(\textbf{r})$ represent the two-dimensional, complex-valued object and let $P(\textbf{r})$ represent the probe while $\textbf{r}$ represents the coordinate vector in the object plane. The object is illuminated at different positions $\textbf{R}_i$ in such a way that the probed areas are overlapping. In the projection approximation \cite{Paganin2006-nt}, the exit field distribution in the object plane $\psi_\mathrm{exit}(\textbf{r},\textbf{R}_i)$ is then given as: 
\begin{equation}
    \psi_\mathrm{exit}(\mathbf{r},\textbf{R}_i) = O(\textbf{r} - \textbf{R}_i) \cdot P(\textbf{r}). 
\end{equation}

The field in the detection plane $\psi_\mathrm{cam}(\mathbf{r^\prime},\textbf{R}_i)$ with coordinate vector $\mathbf{r^\prime}$ can be obtained by choosing an appropriate diffraction model. We have implemented both the angular spectrum method and the Fraunhofer diffraction model, the latter of which conveniently accommodates Fresnel diffraction by absorption of a quadratic phase function into the probe~\cite{Goodman2017-ll}. 
While the angular spectrum method can be used to model both near-field and far-field diffraction, it comes at the cost of computing two fast Fourier transforms (FFT). The Fraunhofer diffraction model, which is based on the far-field approximation, can be performed using a single FFT.
The last step in the forward model is the calculation of the intensity in the detection plane using
\begin{equation}
    I_{\mathrm{pred}}(\mathbf{r^\prime},\textbf{R}_i, \bm{\theta}) = |\psi_\mathrm{cam}(\mathbf{r^\prime},\textbf{R}_i)|^2,
\end{equation}
where $I_{\mathrm{pred}}(\mathbf{r^\prime},\textbf{R}_i, \bm{\theta})$ is the predicted coherent diffraction pattern at the $i$th position given the current best-fit for some parameters $\bm{\theta}$ that describe the model. While the set of parameters $\bm{\theta}$ usually includes amplitude and phase values characterizing the probe field and the object, we will show below that $\bm{\theta}$ can also include other experimental parameters such as the reconstruction distance.
Eventually, the objective of the framework is to minimize the mean squared error (MSE) function given as
\begin{equation}
    \operatorname{MSE}=\frac{1}{N}\sum_{i=1}^N(I_{\mathrm{meas}}(\mathbf{r^\prime},\textbf{R}_i)- I_{\mathrm{pred}}(\mathbf{r^\prime},\textbf{R}_i, \bm{\theta}))^2,
\end{equation}
where $N$ is the total count of scanning points in the chosen illumination scheme, and $I_{\mathrm{meas}}(\mathbf{r^\prime},\textbf{R}_i)$ is the experimentally measured diffraction pattern at scanning \mbox{position $\textbf{R}_i$}. 
As a data standardization method, we normalize the ptychographic data set to the highest measured pixel intensity, a choice that only affects the ensuing tuning of the optimizer hyperparameters.


The key idea of AD-based ptychography is to rely on the optimizer to reconstruct the object and probe.
As computers execute mathematical functions as a sequence of elementary arithmetic operations with known derivatives, the derivative of the objective function with respect to $\bm{\theta}$ is computationally known (by applying the chain rule repeatedly). Therefore, AD enables us to converge to a solution of the ptychographic problem without the need to find an analytical expression for the update function, in contrast to algorithms based on PIE.
In practice, we employ the popular Adam optimizer \cite{Kingma2014-fa} to make changes to $\bm{\theta}$ in each backpropagation epoch.

\section{Simulated data}

\begin{figure}[ht]
\centering
\includegraphics[width=9.6cm]{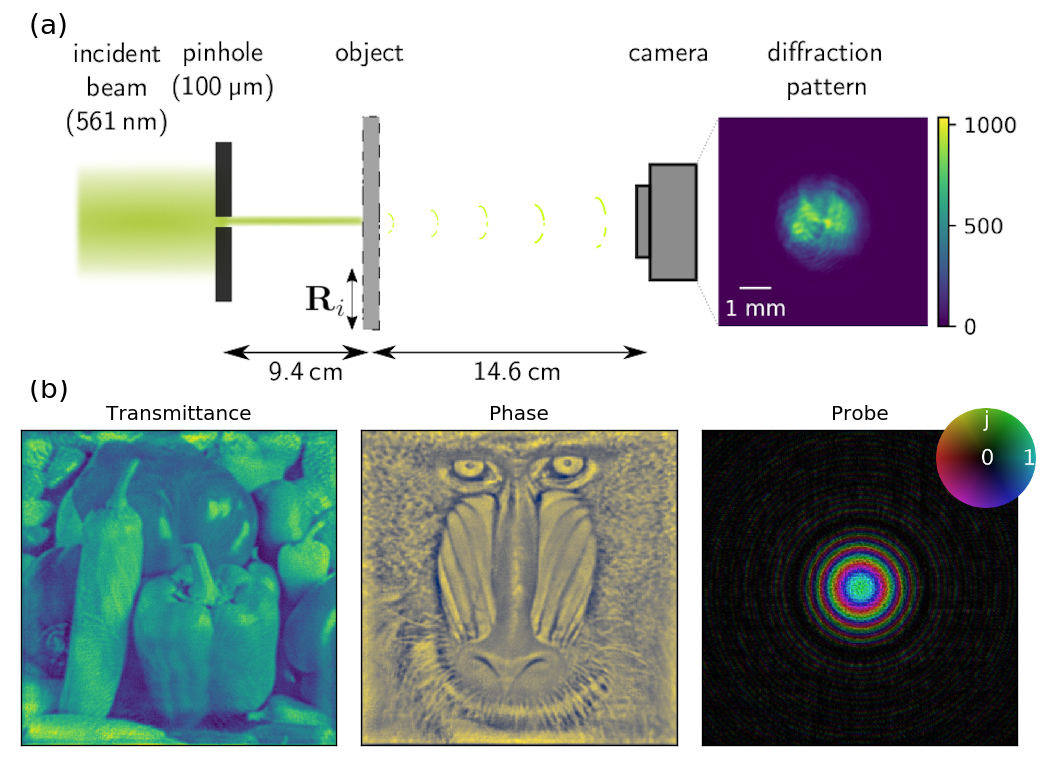}
\caption{(a) A single aperture ptychography setup is simulated to produce synthetic diffraction patterns. The object is laterally displaced with respect to the illumination field. For every position $\bm{R}_i$, the diffraction pattern is collected using a camera downstream of the object. (b) Reconstruction results from synthetic data. The complex-valued reconstruction is decomposed into transmittance and phase contrast. The probe is plotted using a color wheel representing the complex plane. The edge length of every image equals \SI{3.3}{\milli\meter}.}
\label{setup}
\end{figure}

With our goal to quantitatively compare the reconstruction speed and quality for our AD-based framework and mPIE, we simulate diffraction patterns according to the setup shown in \fig{setup}(a).
A ptychographical dataset is synthesized by generating a $512\times512$ complex-valued object generated from two images, respectively defining the transmittance and phase contrast of the object. The pixel size is \SI{6.45}{\micro\meter}. Using coherent illumination at $\lambda=\SI{561}{\nano\meter}$, we shift the object to 80 overlapping locations with a relative overlap of \SI{60}{\percent} as defined and recommended by \mbox{Bunk et al.~\cite{Bunk2008-nq}}. The projection approximation is applied to calculate the exit waves which are then propagated to the detection plane using the angular spectrum method.
To calculate the synthetic diffraction patterns, we then take the absolute squares of the propagated fields and add both (Poissonian) shot noise and (Gaussian) camera read-out noise to the synthetic data. 
The simulated intensities are in the order of a few hundred counts per pixel, and the Gaussian noise is characterized by a standard deviation of $\sigma=10$ counts.
Then, we utilize our optimization framework based on AD to reconstruct the object and probe functions. As shown in \fig{setup}(b), we successfully reconstruct both the complex-valued object and the probe by running the algorithm for 170 epochs (\SI{4}{\minute}).

For a quantitative comparison between our AD-based framework and mPIE, we run them on the same graphics card (NVIDIA GeForce RTX 2070) using the synthetic dataset. To this end, we use the GPU-enabled mPIE implementation from Ref.~\cite{Loetgering2018-wf} and measure both reconstruction quality and speed. It must be stressed that the computational time needed for one iteration in mPIE and for one epoch of our AD-based framework can be different. 
Therefore, to provide a meaningful comparison, we measure the reconstruction speed in time units. In order to quantify reconstruction quality, a useful figure of merit to compare complex functions is the complex correlation coefficient $C$ defined by
\begin{equation}
\label{corr_eq}
    C =  
    \frac{ \sum\limits_{i=1}^{n}{O^*(r_i)  \hat{O}(r_i)} }{ \sqrt{\sum\limits_{i=1}^{n}{|O(r_i)|^2}}  \sqrt{\sum\limits_{i=1}^{n}{|\hat{O}(r_i)|^2}}},
\end{equation}
where $O(\textbf{r})$ is the ground truth, $\hat{O}(\textbf{r})$ is the estimate of the object and $n$ is the total number of pixels per image. Note that the reconstructed object can be shifted in space relative to the ground truth, which is a typical ambiguity in ptychography. We estimate this shift and compensate for it before calculating the correlation coefficient. The absolute value of the coefficient $C$ is a quantitative indicator for how well an estimate resembles the ground truth. The argument of the correlation coefficient can be discarded as it only represents a global phase shift in the reconstruction.

\begin{figure}[ht]
\centering
\includegraphics[width=7.5cm, trim={0.2cm 0.2cm 0.2cm 0.1cm}]{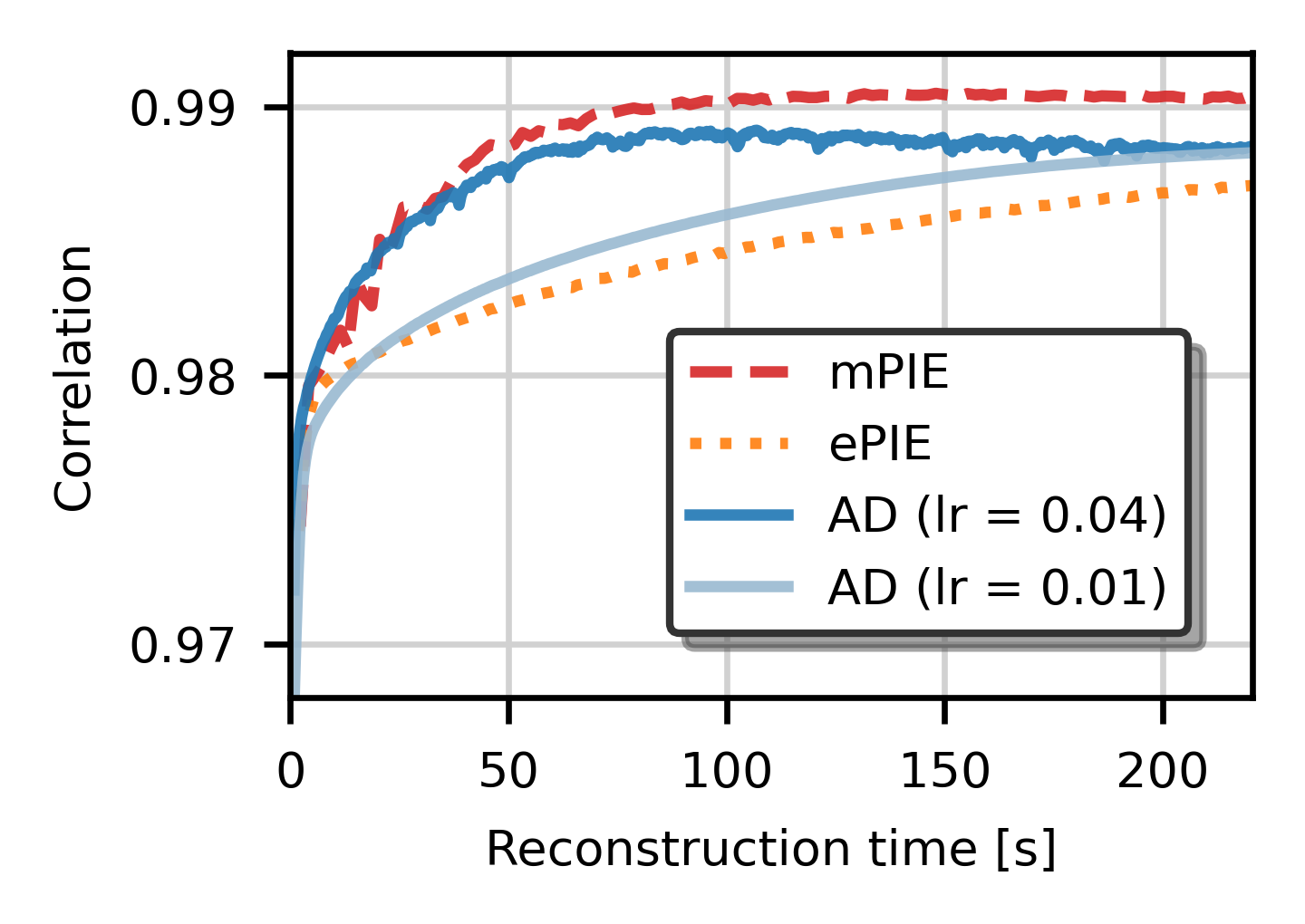}
\caption{Convergence results for reconstructions using mPIE, ePIE and our AD-based framework in simulation. The correlation between the ground truth and the reconstruction estimates is shown as a function of computation time. All algorithms run on the same computer hardware. lr: learning rate for the Adam optimizer.}
\label{correlation}
\end{figure}

The performances of mPIE and of our AD-based framework generally depend on the choice of some hyperparameters. We choose hyperparameters for mPIE according to the values suggested in Ref.~\cite{Maiden2017-tj}. For our AD-based framework, we can control the learning rate of the optimizer, which determines the step size at each epoch while moving toward a minimum of the objective function.
In \fig{correlation}, we show the magnitude of the correlation coefficient as a function of the reconstruction time for different learning rates. We also include ePIE for reference.
Using a learning rate of $0.01$, our AD-based framework converges robustly but is significantly outperformed by mPIE in terms of reconstruction speed.
Using a learning rate of $0.04$, our AD-based framework performs comparably to mPIE. Higher learning rates become numerically unstable and do not converge to a reasonable estimate of the ground truth.
Both algorithms converge rapidly within the first \SI{60}{\second}. 
After this time, we observe a decrease of the reconstruction performance in our AD-based framework due to stronger overfitting to noise in comparison to mPIE. For long reconstruction times of more than \SI{200}{\second}, noise overfitting also occurs in mPIE, but to a significantly smaller extent. 
The influence of noise raises a relevant question for future research on AD-based ptychography. In Ref.~\cite{Thibault2012-mn} and \cite{Godard2012-gr}, different noise models and mitigation strategies such as regularization and maximum-likelihood refinement have already been explored for coherent diffraction imaging. 
It was shown that an adaptive step size strategy could improve the robustness to noise in ptychography~\cite{Zuo2016-ag}. The conceptually similar approach of gradually decreasing the learning rate in our AD-based framework does not produce an immediate solution to the overfitting phenomenon and comes at the cost of introducing another arbitrary hyperparameter, namely the learning rate decay rate.
A comparison for the separate reconstruction qualities for amplitude and phase images is provided in Fig.~S1 in the supplementary document~\ref{more_corr}.

\begin{figure}[htb]
\centering
\includegraphics[width=7.5cm, trim={0.2cm 0.2cm 0.2cm 0.1cm}]{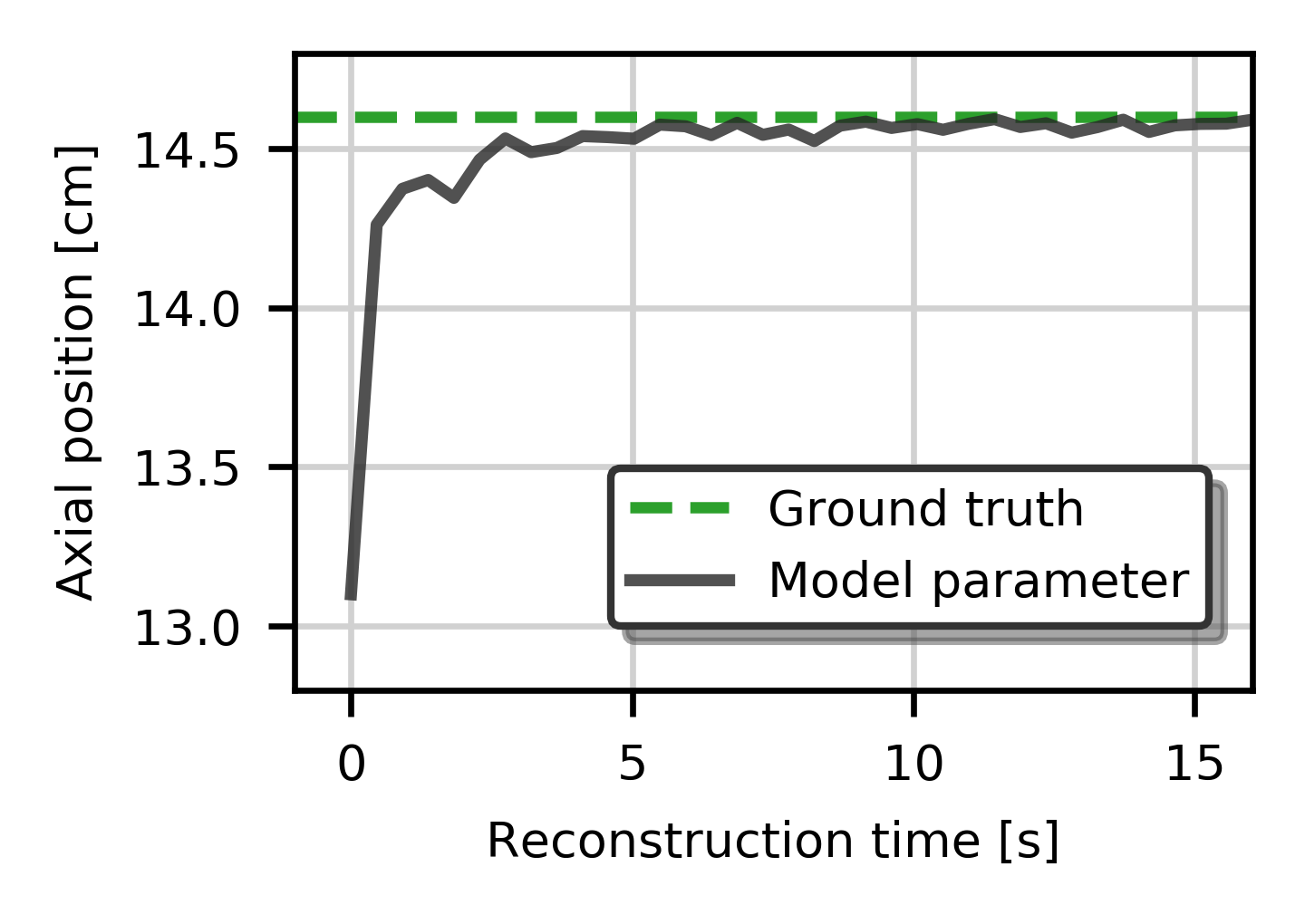}
\caption{Correction of the reconstruction distance (axial distance between camera and object) using our AD-based framework. In the reconstruction, the camera position is defined as trainable and initialized with an error of \SI{1.5}{cm}.}
\label{axialCorrection}
\end{figure}

In conventional ptychography, $\bm{\theta}$ usually comprises the pixel values for $O(\textbf{r})$ and $P(\textbf{r})$ as originally introduced in the ePIE algorithm~\cite{Maiden2009-pn}. However, one important strength of our framework is that we can choose $\bm{\theta}$ relatively freely, e.\,g. by incorporating experimental parameters into the forward model. 
We demonstrate this flexibility by estimating and correcting the axial distance $z$ between the object and detector, similar to the recently shown autofocusing in ptychography using a sharpness metric and an algorithm called zPIE in Ref.~\cite{Loetgering2020-ug}. By using the angular spectrum method in our AD framework, the forward model becomes explicitly dependent on $z$. Therefore, we can easily define $z$ as a trainable parameter of the model in the same way as we defined the pixel values of $O(\textbf{r})$ or $P(\textbf{r})$.
To illustrate this approach, we initialize the reconstruction algorithm with an error of \SI{15}{\milli\meter} in the axial distance $z$. As shown in \fig{axialCorrection}, our AD-based framework is able to find the true value of $z$ with submillimeter precision in approximately \SI{15}{\second} ($35$~epochs), in addition to the pixel values of $O(\textbf{r})$ and $P(\textbf{r})$. Here, note that we only obtain this result by using a precalibrated probe.

\section{Experimental data}

\begin{figure}[t]
\centering
\includegraphics[width=9.5cm]{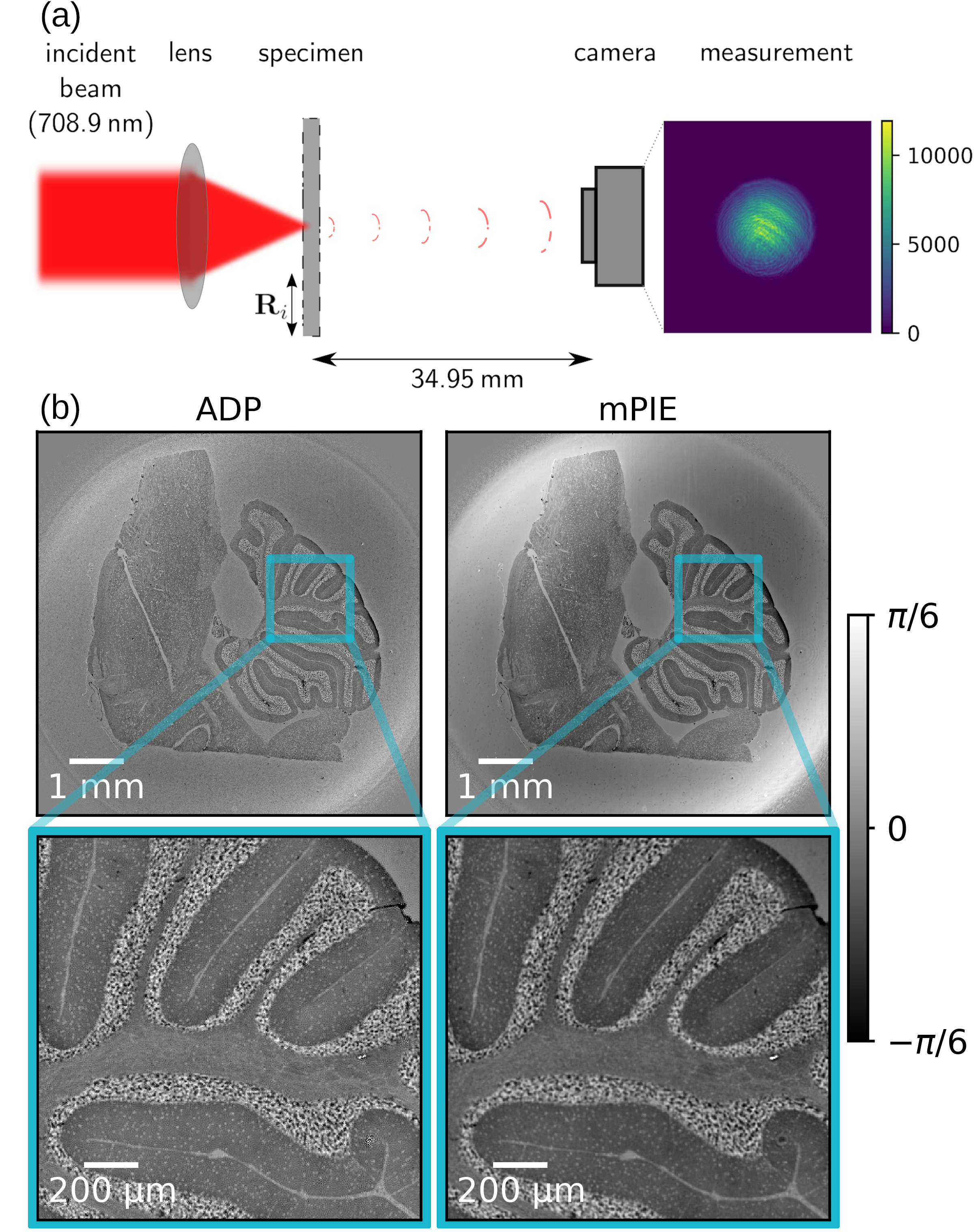}
\caption{(a) Ptychography setup for imaging a histological slice of a mouse cerebellum. A diffraction pattern is measured for every scanning position $\bm{R}_i$ of the sample. (b) Reconstruction results from experimental data. The phase contrast induced by a histological slice of a mouse cerebellum is reconstructed using our AD-based framework and mPIE.}
\label{exp_reconstruction}
\end{figure}

\sisetup{range-phrase=-}
In order to test our AD-based framework on experimental data, we study a biological sample.
We acquire a near-infrared ptychographic scan of a histological slice of a mouse cerebellum.
The sample is illuminated with a focused supercontinuum laser source that is spectrally filtered to a wavelength of $\lambda = \SI{708.9}{nm}$ (filter bandwidth, $\SI{0.6}{nm}$).
The sample is mounted on a XY translation stage (2x Smaract SLC-1770-D-S) with a sample-detector distance of $z = \SI{34.95}{mm}$. 
The XY translation stage is driven by piezoelectric actuators with an accuracy on the position of the order of tens of nanometers, making it unnecessary to employ position correction methods.
A charge-coupled device (CCD) camera (AVT prosilica, $1456\times1456$ pixels with pixel size \SI{4.54}{\micro\meter}) is used to collect 1824 diffraction patterns downstream of the specimen. Reconstructions are carried out in the same way as described before using the synthetic dataset. 
The resulting reconstructions obtained with our AD-based framework and mPIE can be visually judged in  \fig{exp_reconstruction} to be of similar quality and resolution.
A more detailed assessment of the achieved resolution with both algorithms is discussed in Fig.~S4 in the supplementary material.
However, our AD-based framework is more prone to the apparition of high-frequency components in the reconstructed image. This effect, which is possibly due to overfitting to noise, could certainly be mitigated by the use of regularization procedures.

\section{Conclusions}

To conclude, we have evaluated the performances of a ptychographic reconstruction framework based on automatic differentiation. Using numerical simulations, we show that our AD-based framework performs comparably to the state-of-the-art algorithm mPIE in terms of speed and quality of reconstructions. 
Furthermore, we show that the flexibility of the forward model in our AD-based framework can readily be utilized for estimating experimental parameters in addition to the pixel values of the probe and object. As an example, we have successfully corrected the reconstruction distance.
Lastly, we have experimentally demonstrated that our framework faithfully reconstructs a biological specimen with a large space-bandwidth product.
We believe that the presented results are important for establishing optimization frameworks based on AD as viable methods within the field of coherent diffraction imaging. Moreover, we can expect AD-based techniques to further improve thanks to the fast-paced progress in machine-learning toolboxes like TensorFlow and in computer hardware like application-specific integrated circuits (\mbox{e.\,g., tensor processing units \cite{Jouppi2017-dp}}).

\bigskip
\paragraph*{Funding.}
Netherlands Organization for Scientific Research NWO (Vici 68047618 and Perspective P16-08).
\bigskip

\paragraph*{Acknowledgements.}
The authors thank C. de Kok for technical support and K. Schilling for providing the biological specimen. This work was supported by the Netherlands Organization for Scientific Research NWO (Vici 68047618 and Perspective P16-08).
\bigskip

\paragraph*{Disclosures.}
The authors declare no conflicts of interest.
\bigskip

\noindent See \href{https://doi.org/10.6084/m9.figshare.13008155}{Supplement 1} for supporting code and design files.


\bibliography{references}


\bigskip



\onecolumngrid
\clearpage
\beginsupplement
\begin{center}
\textbf{\large Efficient and flexible approach to ptychography using an optimization framework based on automatic differentiation\\ \vspace{0.5cm} Supplementary information}

\bigskip
Jacob Seifert,$^1$, Dorian Bouchet,$^1$ Lars Loetgering,$^2$ and Allard P. Mosk$^1$\\ \vspace{0.15cm}
\textit{\small $^\mathit{1}$Nanophotonics, Debye Institute for Nanomaterials Science,\\ Utrecht University, P.O. Box 80000, 3508 TA Utrecht, the Netherlands}\\
\textit{\small $^\mathit{2}$Advanced Research Center for Nanolithography, Science Park 106, 1098 XG Amsterdam, The Netherlands}
\end{center}
\vspace{1cm}

This document provides supplementary information to ``Efficient and flexible approach to ptychography using an optimization framework based on automatic differentiation''.

\section{Amplitude and phase correlations}

\begin{figure}[htb]%
    \centering
    \includegraphics[width=6.2cm]{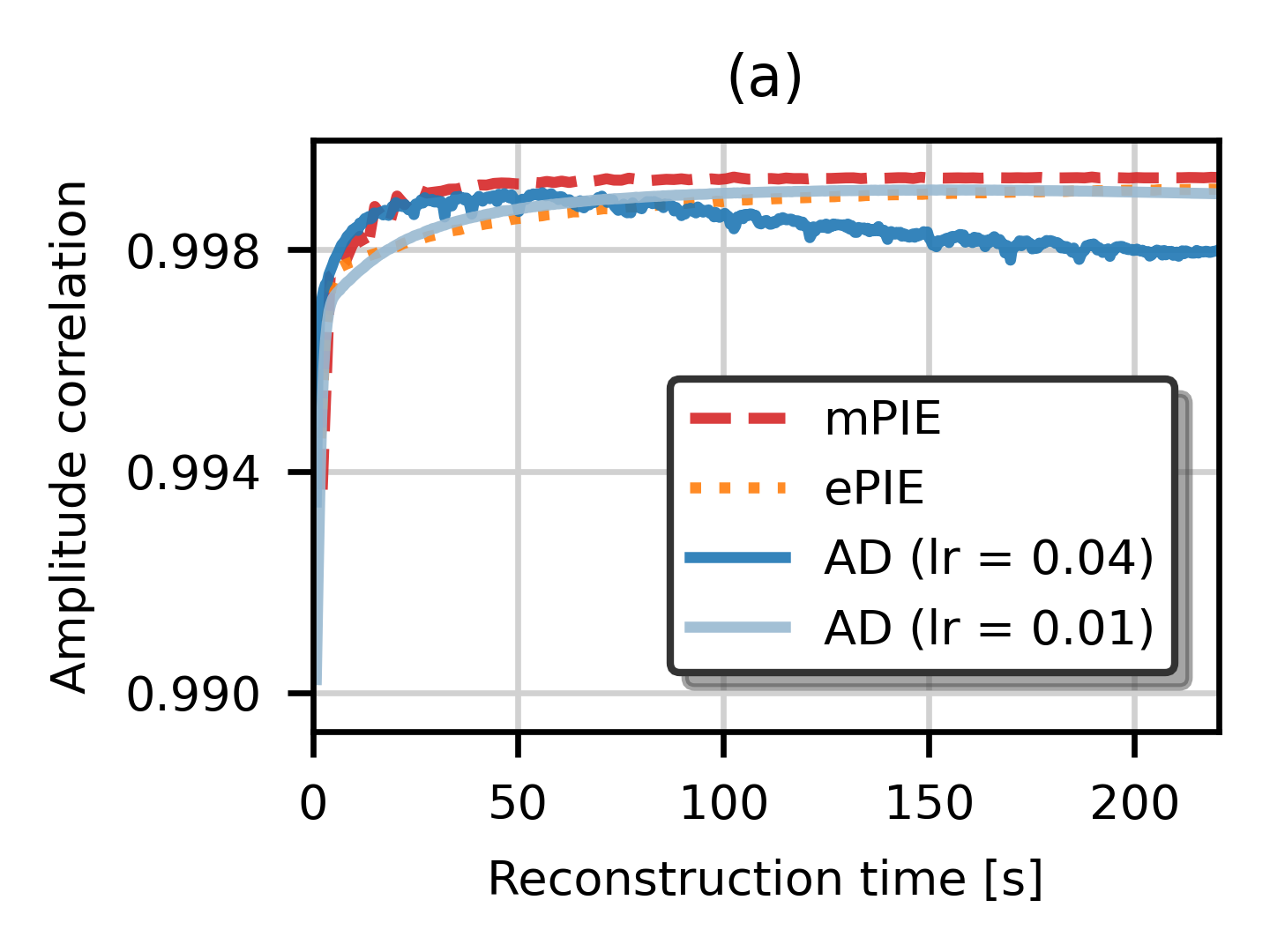} %
    \qquad
    \includegraphics[width=6.3cm]{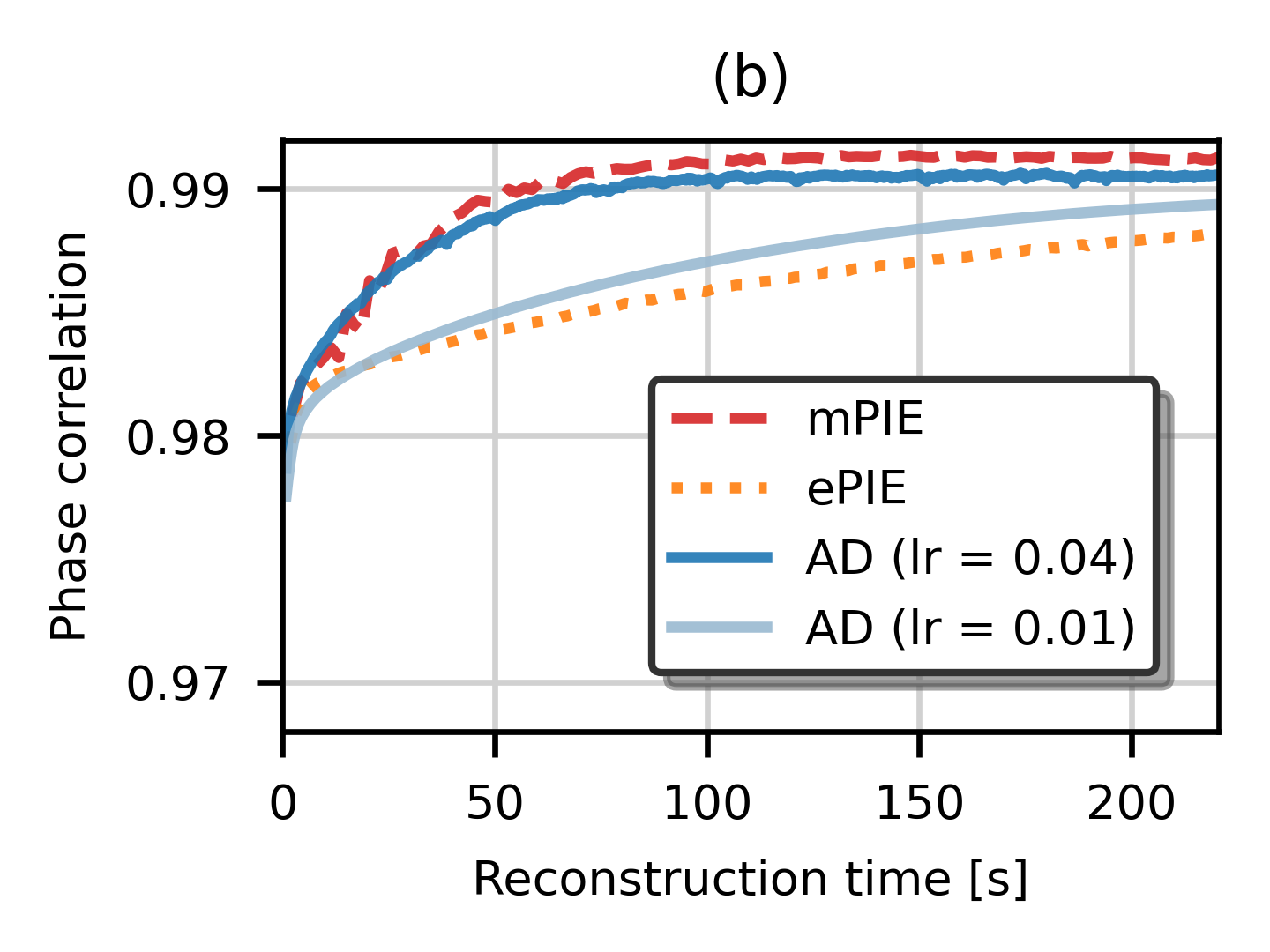}%
    \caption{\textbf{Fig. S1.} Convergence results for reconstructions using mPIE, ePIE and our AD-based framework in simulation for (a) amplitude images and (b) phase images. The correlations between the ground truth and the reconstruction estimates are shown as a function of computation time. All algorithms run on the same computer hardware. lr: learning rate for the Adam optimizer.}%
    \label{more_corr}%
\end{figure}

Correlating the ground truth with the reconstructed complex-valued object using equation~(4) provides an accurate measure of the overall reconstruction performance. To gain knowledge about the reconstruction performance of the amplitude and phase images independently, it is useful to correlate them to the ground truth separately. 
Since the phase values in ptychography are cyclic variables and not absolute, one has to avoid spurious influences from phase wrapping and global phase shift when comparing two phase images. 
This is achieved by computing the magnitude of the complex correlation of $\exp[i\varphi(\mathbf{r})]$, where $\varphi(\mathbf{r}) = \arg[O(\mathbf{r})]$ represents the phase image.
The results are shown in Fig.~\ref{more_corr}.

\section{Diffraction intensities and scanning trajectories}

\begin{figure}[t]%
    \centering
    \includegraphics[width=12.0cm, trim=25 25 15 20, clip]{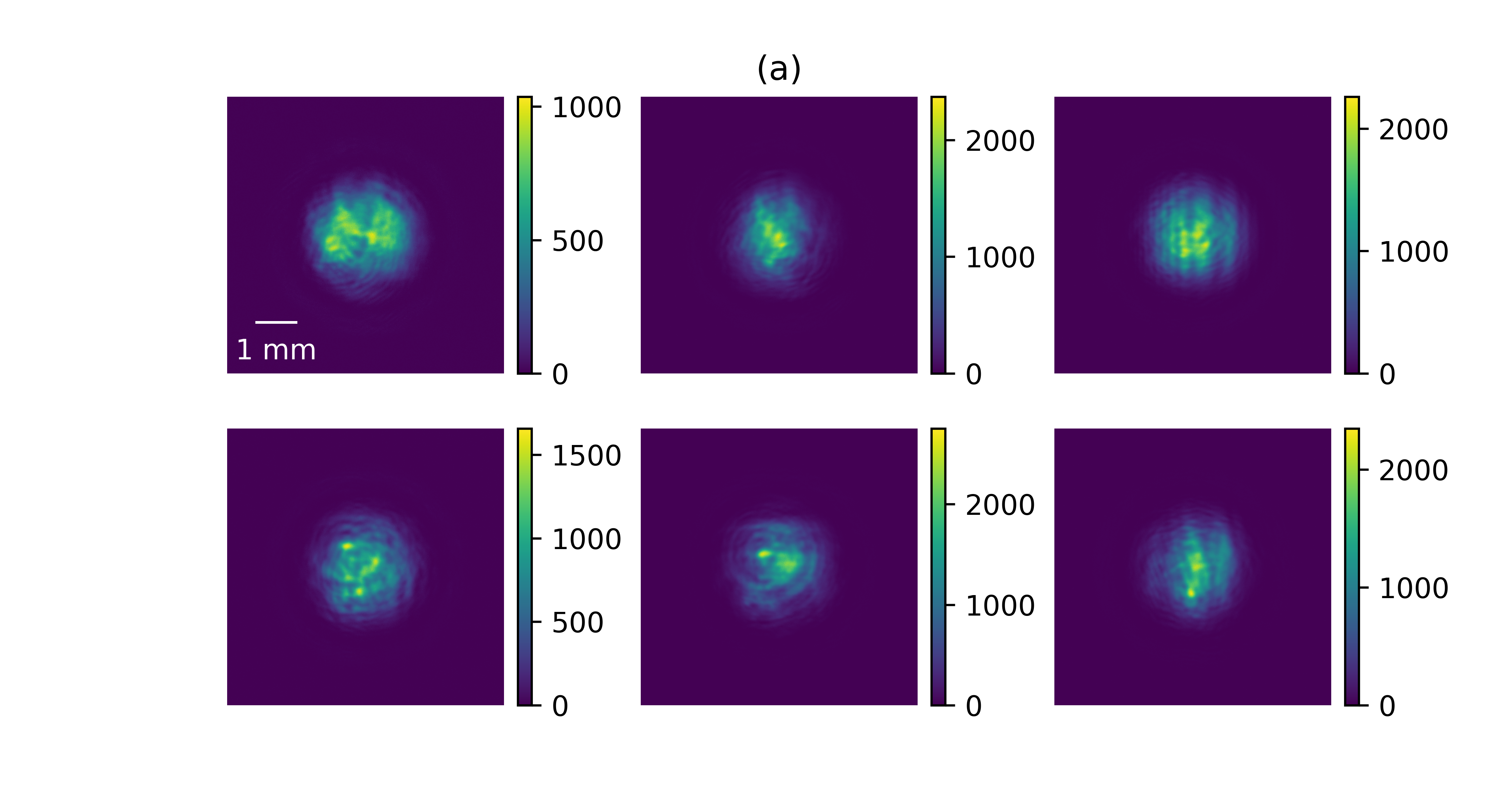} %
    \qquad
    \includegraphics[width=12.0cm, trim=25 25 15 20, clip]{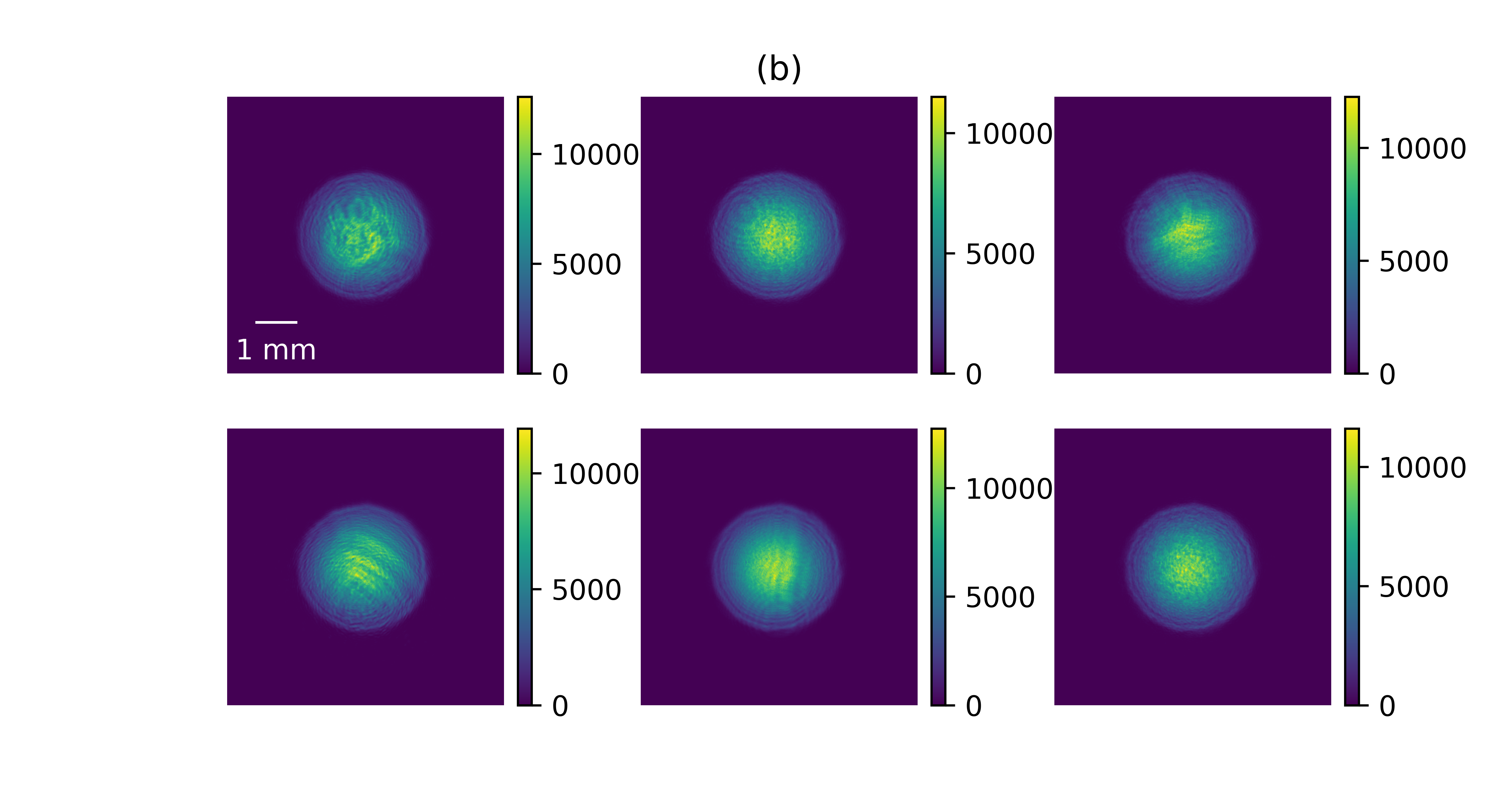}%
    \caption{\textbf{Fig. S2.} Random selection of diffraction intensities used for phase retrieval. (a) Synthetic data and (b) experimental data. The colorbars denote the counts of the photoelectric detector.}
    \label{diff}%
\end{figure}

\begin{figure}[ht]%
    \centering
    \includegraphics[width=5.7cm, trim=0 0 0 12, clip]{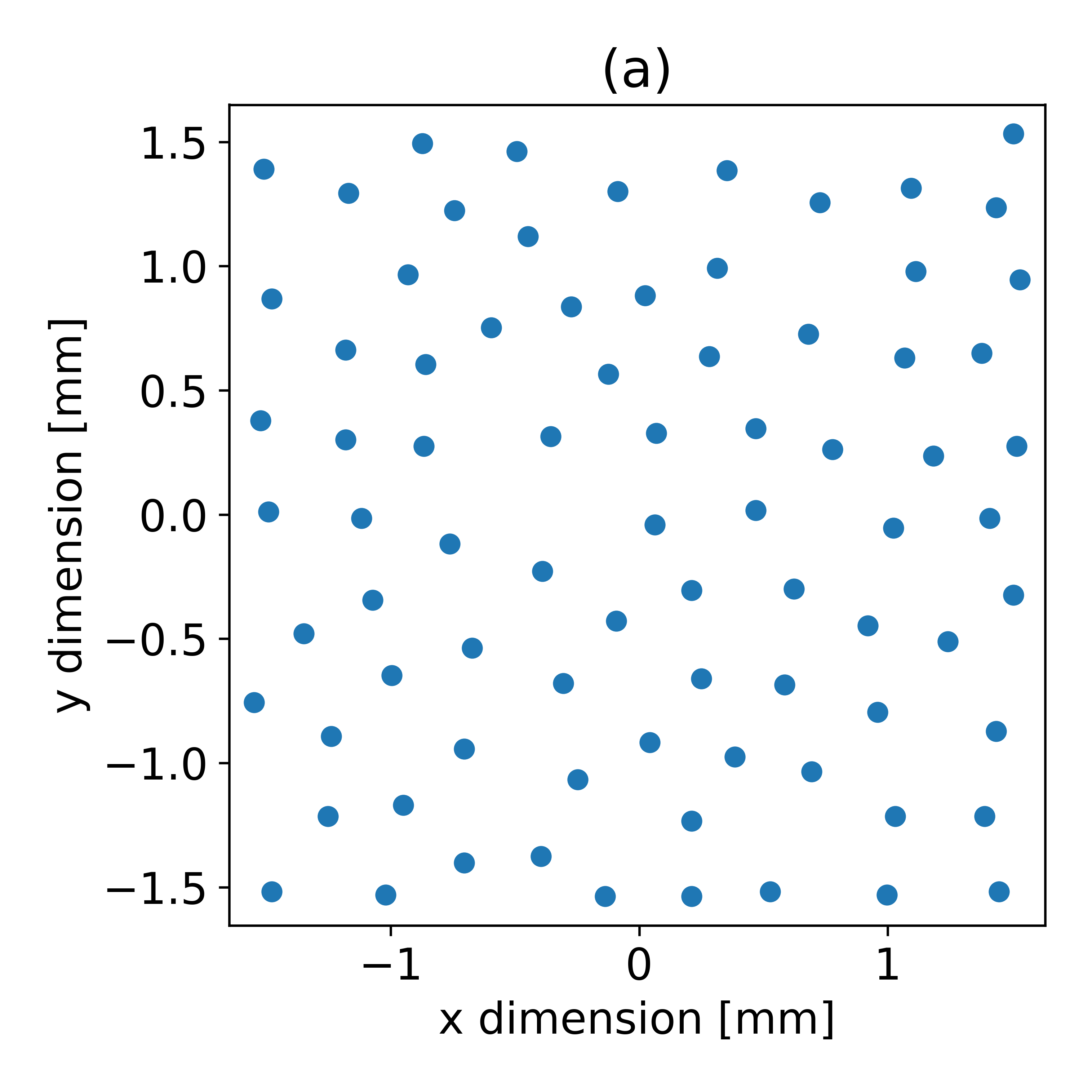} %
    \qquad
    \includegraphics[width=5.7cm, trim=0 0 0 12, clip]{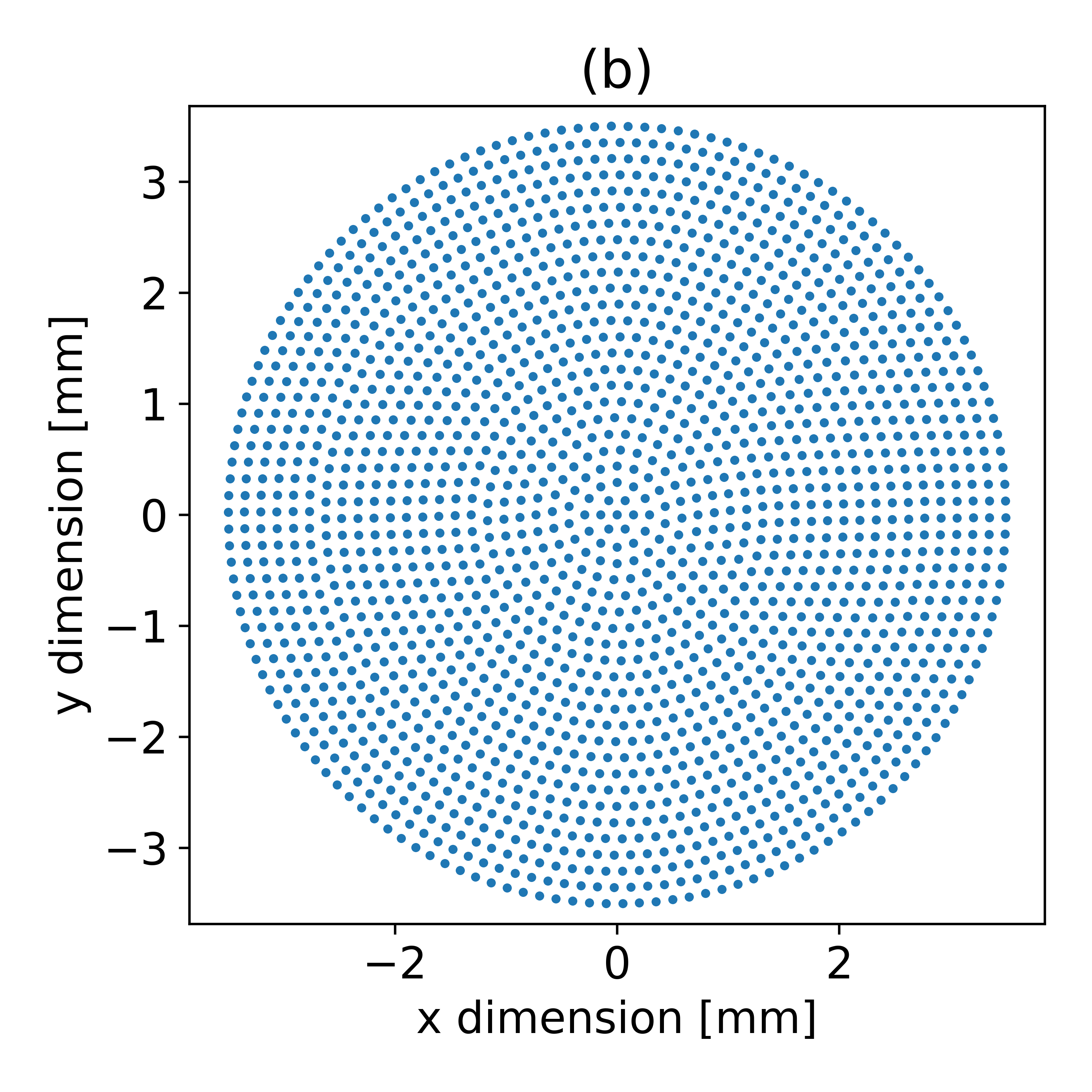} %
    \caption{\textbf{Fig. S3.} Scanning trajectories for the probe positions used to illuminate the object. (a) Synthetic data with a total of 80 positions and (b) experimental data with a total of 1824 positions.}%
    \label{pos}%
\end{figure}

In Fig.~\ref{diff}, a selection of randomly selected diffraction intensities for both the synthetic and experimental ptychography datasets are shown. 
The diffraction intensities are shown after the application of Poisson noise and readout noise, and are sampled with a dynamic range of 12 bits.
Furthermore, the scanning trajectories for the probe positions are visualized in Fig.~\ref{pos}. To avoid the raster grid pathology in ptychography, we utilize trajectories following a concentric pattern for the experiment and Poisson disk sampling for the synthetic data~\cite{Huang2014-yo, Bridson2007-mk}. 

\newpage

\section{Reconstruction of a resolution test sample}

\begin{figure}[hb]%
    \centering
    \includegraphics[width=9.5cm]{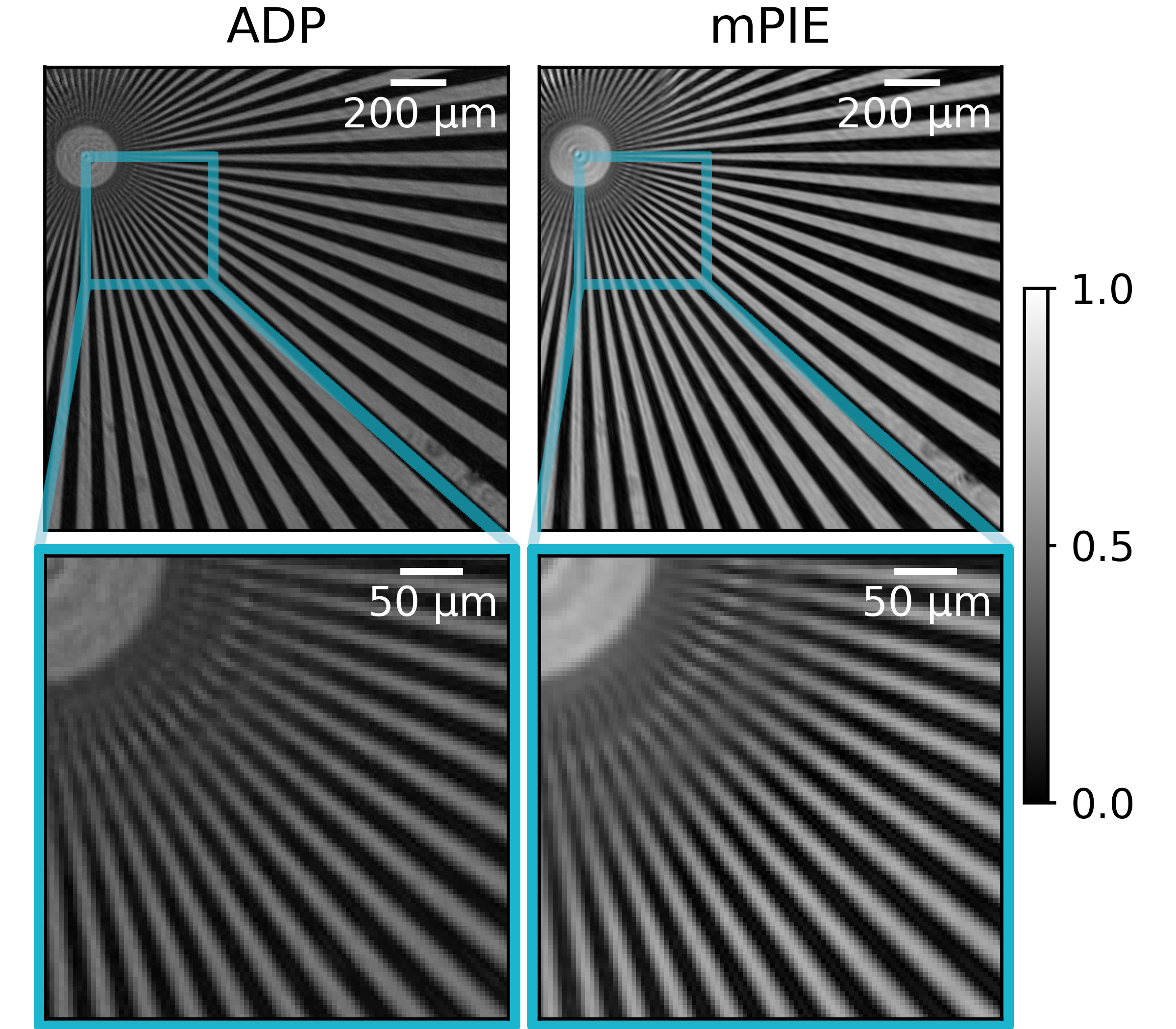} 
    \caption{\textbf{Fig. S4.} Reconstruction results from experimental data. The transmittance contrast of a sector star is reconstructed using our AD-based framework and mPIE.}%
    \label{sector_star}%
\end{figure}

A quantitative assessment of the resolution in a biological specimen is difficult. Fourier ring correlation (FRC) is an unbiased and general method to measure the degree of correlation of the two images at different spatial frequencies. 
Despite it being promising, we have observed that FRC can produce spurious correlations in ptychography, making it hard to interpret the results. 
Therefore, we are falling back to the reliable approach of imaging a sector star test sample. 
Subsequently, we can compare the achieved resolutions from reconstructing the data set using both mPIE and our AD-based framework.
The experimental setup is identical to the one used for the reconstruction of the mouse cerebellum and shown in Fig.~5 in the main document. Only the object-camera distance has been changed from \SI{34.95}{mm} to \SI{46.87}{mm}.
The reconstructed transmittances are shown in Fig.~\ref{sector_star}. The inner center circle of the sector star has a diameter of \SI{200}{\micro\meter} and a total of 72 lines, resulting in a density of 115 lines per mm at the center circle or a line width of \SI{8.7}{\micro\meter}. It should be noted that both reconstructions do not entirely resolve the very finest details of the sector star, indicating an achieved effective resolution of roughly \SI{10}{\micro\meter}. No significant difference between mPIE and our AD-based reconstruction is present. Note that the diffration limit of the optical system with the given object-camera distance and detector size is approximately \SI{3}{\micro\meter}.

\end{document}